\documentclass[pra,showpacs,twocolumn,superscriptaddress,floatfix]{revtex4}
\usepackage{times,bbm,hyperref,amssymb,amsmath,graphicx}
\newcommand{\qfi}{{\cal H}}

\newcommand{\hc}{\mathrm{h.c.}}

\newcommand{\LL}{{\mathcal L}}
\newcommand{\1}{\leavevmode{\rm 1\ifmmode\mkern  -4.8mu\else\kern -.3em\fi I}}
\begin{document}
\title{Quantum criticality as a resource for quantum estimation}
\author{Paolo Zanardi}
\email{zanardi@usc.edu}
\affiliation{Department of Physics and Astronomy, University of Southern
California, Los Angeles, CA 90089-0484 (USA) }
\affiliation{Institute for Scientific Interchange, I-10133 Torino, Italia}
\author{Matteo G. A. Paris}
\email{matteo.paris@fisica.unimi.it}
\affiliation{Institute for Scientific Interchange, I-10133 Torino, Italia}
\affiliation{Dipartimento di Fisica dell'Universit\`a di Milano, I-20133, Italia.}
\affiliation{CNISM, Udr Milano Universit\`a, I-20133, Milano, Italia.}
\author{Lorenzo Campos Venuti}\email{campos@isi.it}
\affiliation{Institute for Scientific Interchange, I-10133 Torino, Italia}
\date{\today}
\begin{abstract}
We address quantum critical systems as a resource in quantum estimation
and derive the ultimate quantum limits to the precision of any estimator
of the coupling parameters. In particular, if $L$ denotes the size of a
system and $\lambda$ is the relevant coupling parameters driving a
quantum phase transition, we show that a precision improvement of order
$1/{L}$ may be achieved in the estimation of $\lambda$ at the critical
point compared to the non-critical case.  We show that analogue results
hold for temperature estimation in classical phase transitions. Results
are illustrated by means of a specific example involving a fermion
tight-binding model with pair creation (BCS model).
\end{abstract} 
\maketitle
\section{Introduction}
It is often the case that a quantity of interest is not directly
accessible, either in principle or due to experimental
impediments.  In these situations one should resort to indirect
measurements, inferring the value of the quantity of interest by
inspecting a given probe. This is basically a parameter
estimation problem whose solution may be found using tools from
classical estimation theory \cite{clet} or, when quantum systems are
involved, from its quantum counterpart \cite{qet}.
Indeed, quantum estimation theory has been successfully applied
to find optimal measurements and, in turn, to evaluate the
corresponding lower bounds on precision, for the estimation of
parameters imposed by both unitary and nonunitary transformations.
These include single-mode phase-shift \cite{Hol79,Dar98,Mon06},
displacement \cite{Hel74}, squeezing \cite{Mil94,Chi06} as well
as depolarizing \cite{Fuj01} or amplitude-damping \cite{Zhe06}
channels in finite-dimensional system, lossy channel in
infinite-dimensional ones \cite{Dau06,QLoss}, and the position of
a single photon \cite{frie07}. Here we focus on the estimation of the parameters 
driving the dynamics of an interacting many-body  quantum system {\em i.e.}, 
the coupling constants defining the system's  Hamiltonian and 
temperature. It is a generic fact that when those are changed the
system is driven into different phases. When this happens at zero
temperature one says that a {\em quantum phase transition} (QPT) 
as occurred \cite{sachdev}.  Different phases of a system are, 
by definition, characterized by radically different physical 
properties, e. g. the expectation value of some distinguished
observable (order parameter). This in turn implies that the
corresponding quantum states have to be statistically distinguishable
more effectively than states belonging to the same phase. 
It is a non trivial result that the degree of
statistical distinguishability is quantified by the Hilbert-space (pure
states) distance \cite{Woo} or by the density operator distance (mixed
states) \cite{Bra94}.  This amount to say that the Hilbert-space
geometry is an {\em information-space geometry} \cite{bro9x,flam,mp08}.  
One is then naturally
led to consider the distance functions between infinitesimally close
quantum states obtained by an infinitesimal change of the parameters
defining the system's Hamiltonian.  Boundaries between different phases
are then tentatively identified with the set of points where this small
change of parameters gives rise to a major change in the distance i.e.,
major enhancement of the statistical distinguishability.  This is
precisely the strategy advocated in
\cite{za-pa,zhou,za-co-gio,co-gio-za,co-ion-za,DG-qpt} 
in the so-called metric (or fidelity) approach to critical phenomena.
\par
The purpose of this paper is to establish the following result: {\em
quantum criticality represent a resource for the quantum
estimation of Hamiltonian parameters}. Indeed, by exploiting the
geometrical theory of quantum estimation, we will show that the
accuracy of the estimation of coupling constants and field
strengths at the critical points is greatly enhanced with respect
to the non-critical ones.  We will also discuss ultimate limits
imposed by quantum mechanics to this scheme of parameter
estimation.  These results are, under several points of view,
analogue of those showing that 
entanglement is a useful metrological resource \cite{em,ac,gml}.
In particular, if $L$ denotes the size of a system and $\lambda$
is the relevant coupling parameters driving a quantum phase
transition, we will show that a relative improvement of order $L$ 
may be achieved in the estimation of $\lambda$ at the critical 
point. 
\par
The paper is structured as follows: In Section \ref{s:geoe} we 
provide a brief introduction to the geometrical theory of quantum 
estimation whereas in Section \ref{s:geoq} we review the metric 
approach to quantum criticality. In Section \ref{s:crtr} we 
show how quantum critical systems represent a resource for 
quantum estimation and derive general results about ultimate quantum 
limits to precision. In Section \ref{s:meas} we illustrate properties
of the optimal measurements, also for systems at finite temperature, 
and the connection with the optimal measurement for state
discrimination. In Section \ref{s:ebcs} we illustrate our general results 
by means of a specific example involving a fermion tight-binding model 
with pair creation. Section \ref{s:out} closes the paper with some
concluding remarks. 
\section{Geometry of Quantum Estimation theory}
\label{s:geoe}
The solution of a parameter estimation problem amounts to find
an estimator, {\em i.e} a mapping $\hat\lambda=\hat\lambda
(x_1,x_2,...)$ from the set $\chi$ of measurement outcomes into
the space of parameters.  Optimal estimators in classical
estimation theory are those saturating the Cramer-Rao inequality,
$$V_\lambda[\hat\lambda] \geq F^{-1} (\lambda)$$ which poses
a lower bound on the mean square error $V_{\lambda}
[\hat\lambda]_{jk} = E_{\lambda} [(\hat\lambda -
\lambda)_j(\hat\lambda-\lambda)_k]$ in terms of the Fisher
information $$F(\lambda) = \int_\chi d\hat\lambda(x)\,
p(\hat\lambda|\lambda) [\partial_\lambda \log
p(\hat\lambda|\lambda) ]^2\:.$$  Of course for unbiased estimators,
as those we will deal with, the mean square error is equal to the
covariance matrix $V_{\lambda} [\hat\lambda]_{jk} = E_{\lambda}
[\hat\lambda_j\hat\lambda_k]  - E_{\lambda} [\hat\lambda_j]
E_{\lambda}[\hat\lambda_k]\:.$
\par
When quantum systems are involved any estimation problem may be stated
by considering a family of quantum states $\varrho(\lambda)$ which are
defined on a given Hilbert space ${\cal H}$ and labeled by  a parameter
$\lambda$ living on a $d$-dimensional manifold  ${\cal M}$, with the
mapping $\lambda \mapsto \varrho(\lambda )$ providing a coordinate
system.  This is sometimes referred to as a quantum statistical model.
In turn, a quantum estimator $\hat\lambda$ for $\lambda$ is 
a selfadjoint operator, which
describe a quantum measurement followed by any classical data processing
performed on the outcomes.
As in the classical case, the goal of a quantum inference process is to
find the optimal estimator. The ultimate precision attainable by quantum
measurements in inferring the value of a parameter or a set of
parameters is expressed by the quantum Cramer-Rao (QCR) theorem
\cite{Hel6X,Bra94} which sets a lower bound for the mean square 
error of the quantum Fisher information.
The QCR bound is independent on the measurement and very much based on
the geometrical structure of the set of the involved quantum states.
The {\em symmetric logarithmic derivative} $\LL (\lambda)$ (SLD)
is implicitly defined as the (set) Hermitian operator(s) satisfying the
equation(s) 
$$\partial_\mu \varrho(\lambda ) 
= \frac12 \left\{ \varrho (\lambda) \LL_\mu (\lambda) +
\LL_\mu (\lambda) \varrho (\lambda) \right\}\:,$$ 
where $\partial_\mu:=\partial/\partial_\mu,\,(\mu=1,\ldots,d)$.
From the e above equation, when $\varrho_j+\varrho_k>0,$ one obtains 
$$\langle\varphi_j|\LL_\mu(\lambda)|\varphi_k\rangle=
2 \langle\varphi_j|
\partial_\mu\varrho(\lambda)|\varphi_k\rangle/(\varrho_j+\varrho_k)\,,$$ 
where we have  used  the spectral resolution
$\varrho (\lambda) =\sum_k \varrho_k\, |\varphi_k\rangle\langle
\varphi_k|$.
The quantum Fisher information $\qfi (\lambda )$ (QFI) 
is a matrix defined as follows
$$\qfi_{\mu\nu}(\lambda) = 
\hbox{Tr}\left[\varrho(\lambda )\,\frac12 
\left\{\LL_\mu(\lambda )\LL_\nu(\lambda ) +
\LL_\nu(\lambda )\LL_\mu(\lambda )\right\}
\right]\:.$$ 
The QFI is symmetric, real and positive semidefinite, {\em i.e.} 
represents a metric for the manifold underlying the quantum
statistical model \cite{bro9x,mp08}.
The QCR theorem states that the mean square error of any quantum estimator
is bounded by the inverse of the quantum Fisher information. In formula
\begin{equation}
V_{\lambda}  [\hat\lambda] \geq 
\qfi^{-1}(\lambda)
\label{QCR}\:.
\end{equation}
QCR is an ultimate bound: it does depend on the geometrical
structure of the quantum statistical model and does not depend on
the measurement. Notice that due to noncommutativity of quantum
mechanics the bound may be not attainable, as it is the case for
several multiparameter models.  When one has at disposal $M$
identical copies of the state $\varrho_{\lambda}$ the QCR bound
reads $V_{\lambda} [\hat\lambda]  \geq 1/M \qfi(\lambda )$, which
is easily derived upon exploiting the additivity of the quantum
Fisher information.  
A relevant remark \cite{Hel6X,Bra94} is that the SLD itself
represents an optimal measurement
 and the corresponding Fisher information is equal 
to the QFI. 
\section{Geometry of Quantum Phase Transitions}
\label{s:geoq}
Now we turn to illustrate the achievements of the metric approach
to QPTs that are relevant to the problem investigated in this
paper.  The crucial point is that this
approach, being based on the state-space
geometry is universally applicable to any statistical system and
does not require any preliminary understanding of the structure
of the different phases e.g., symmetry breaking patterns, order
parameters. In principle not even the system's Hamiltonian has to
be known as long as the relevant states are (see e.g., the
analysis of matrix-product state QPTs \cite{co-ion-za}). Of
course the main conceptual as well as technical obstacle one has
to overcome in order to get this simple strategy at work is
provided by the fact that the relevant phenomena are
intrinsically thermodynamical limit ones.  More technically, one
considers
the set of Gibbs thermal states $\varrho_\beta(\lambda):= Z^{-1}
e^{-\beta H(\lambda)}, \,(Z:=\hbox{Tr}[ e^{-\beta H(\lambda)}])$
associated with a parametric family of Hamiltonians
$\{H(\lambda)\}_{\lambda\in{\cal M}}.$ Physically the $\lambda$'s are to
be thought of as coupling constants strengths and external fields
defining the many-body Hamiltonian $H(\lambda)$. The systems one is
interested in are characterized by the fact that, in the thermodynamical
limit, they feature a zero-temperature i.e., quantum, phase transitions
(QPT) for critical values $\lambda_c$ \cite{sachdev}.  We now consider
the Bures metric $ds_B^2 = \sum_{\mu\nu} g_{\mu\nu}\,
d\lambda_\mu d\lambda_\nu$ 
over the manifold of density matrices where
\cite{bures-formula}
\begin{equation}
g_{\mu\nu}=
\frac{1}{2}\sum_{jk}\frac{\langle
\varphi_j|\partial_\mu\varrho|\varphi_k\rangle\langle
\varphi_k|\partial_\nu\varrho|\varphi_j\rangle}
{\varrho_k+\varrho_j}
\label{bures1}
\end{equation}
The key result of the metric (or fidelity) approach to QPTs is that the
set of critical parameters can be identified and analyzed in terms of
the scaling and finite-size scaling behavior of the metric
(\ref{bures1}).  More precisely the metric has the following properties:
{\bf i)}In the thermodynamical limit and in the neighborhood of the
critical values $\lambda_c$ the zero-temperature metric has the following
scaling behavior 
$ds^2_B\sim L^d |\lambda-\lambda_c|^{-\nu \Delta_g},$ 
where $L$  is the system size, $d$ the spatial dimensionality, $\nu$ 
the correlation length exponent ($\xi\sim |\lambda-\lambda_c|^{-\nu}$)
and $\Delta_g =2 \zeta +d -2
\Delta_V$. Here $\zeta$ is the dynamical exponent and $\Delta_V$ is
the scaling dimension of the operator coupled to $\lambda$. 
{\bf ii)} 
At the critical points, or more generally in the critical region
defined by $L\ll\xi$, the finite-size
scaling is as follows 
$ds_B^2\sim L^{d+\Delta_g}$. 
The main point is that {\em for a wide class of QPTs $\Delta_g$ 
can be greater than zero,
whereas at the regular points the scaling is always extensive
i.e., $ds_B^2\sim L^d$}. To be precise superextensive behavior
requires that the perturbation be
sufficiently relevant.  
The superextensive behavior gives rise, for $L\rightarrow\infty$, to a peak
(drop) of the metric (fidelity) that allows one to identify the boundaries
between the different phases. 
Moreover it can be proven \cite{lor} that, for
local Hamiltonians, superextensive behavior of any of the metric elements is
a sufficient condition for gaplessness i.e., criticality.  
On the other hand criticality its {\em not} a sufficient condition for such
a super-extensive scaling. There are indeed QPTs driven by local operators
not sufficiently relevant (renormalization group sense) where no  peak
in the metric is observed at the critical point \cite{lor}.
{\bf iii)} When the
temperature is turned on one can still see the signatures of quantum
criticality. This is done by studying the scaling, as a function of the
temperature, of the elements of the metric (\ref{bures1}) or fidelity
\cite{zhong-guo}.  In particular when the temperature is small but bigger than
the system's energy gap one has 
$ds_B^2\sim T^{-\beta}$, $(\beta>0)$. 
When one sits, in the parameter space, at the critical point 
this result can be extended all the way down to zero temperature
giving rise to a divergent behavior 
that matches $L |\lambda-\lambda_c|^{-\nu\Delta_g}$. 
Remarkably, crossovers between semiclassical and quantum 
critical regions in the $(T,\lambda)$ plane can be identified 
by studying the largest eigenvalue of the metric or its curvature \cite{ZCG}.
\section{Criticality as a resource}
\label{s:crtr}
To the aims of this paper it is crucial to notice that
the QFI  is proportional to the Bures metric (\ref{bures1}).
Indeed, by evaluating the trace defining the QFI in the 
eigenbasis of  $\varrho (\lambda)$
one readily finds \cite{Bra94}
$g_{\mu\nu} = \frac14 \qfi_{\mu\nu}$. 
This simple remark, along with the results of the metric approach to
criticality summarized in the previous section, immediately lead to the
main conclusion of this paper: {\em  the estimation of a physical
quantity driving  a quantum phase transitions is dramatically enhanced
at the quantum critical point}. 
It is important to notice that the Hamiltonian dependence on the
quantity to be estimated can be even an 
indirect one. More precisely, suppose that $H$ depends on  a set
of coupling constants $\lambda$ and those 
in turn depend on the unknown (to be estimated) quantities
$\lambda^\prime$ i.e., 
$\lambda=f(\lambda^\prime),$ through a known 
function $f;$ we assume also that $f$ is smooth and that its
derivatives are bounded and system's size 
independent.  From the tensor nature of the Bures metric i.e.,
$g_{\mu\nu}^\prime=g_{\alpha\beta}
(\partial \lambda_\alpha/\partial\lambda_\mu^\prime) (\partial
\lambda_\beta/\partial\lambda_\nu^\prime),$
one has that the QFI associated to the 
$\lambda^\prime$
has the same dependence on the system's size of the QFI
associated to the $\lambda$'s and the same divergencies in the
thermodynamical limit. From the QCR bound (\ref{QCR}) then it
follows that  if one is able to engineer a system such that the
coupling constants $\lambda$ defining its Hamiltonian (featuring
QPTs) depends on the unknown quantities $\lambda^\prime$ in the
way outlined above then the  $\lambda^\prime$'s can be estimated
with a greater efficiency in the points corresponding, to the
QPTs. For example if $\lambda=\mu_0-\lambda^\prime,$ then
$\lambda^\prime$ can be effectively estimated around
$\lambda^\prime=\mu_0-\lambda_c$ ($\mu_0$ is
assumed to be {\em exactly} known).
\par
In order to quantitatively asses the improvement in the  
estimation accuracy, let us focus on the single parameter case
$\lambda\in {\bf{R}}.$ In this case the QCR  reads
$V_\lambda[\hat{\lambda}]\ge (4 g_{11})^{-1}.$ From the results
summarized in the previous section one easily sees that
{\bf i)} In the neighborhood of  critical point and at zero
temperature the optimal covariance $V_\lambda[\hat{\lambda}]$
scales like  $|\lambda-\lambda_c|^{\nu\Delta_g}$. This is a remarkable
fact for at least two reasons. On the one hand, it means that the
covariance itself scales as the parameter and thus divergence of the
QFI are allowed \cite{QLoss}. On the other hand, for those QPTs such
that $\Delta_g>0$, the covariance can be then pushed all
the way down to zero by getting closer and closer to the quantum
critical point.  
{\bf ii)} The covariance of $\hat{\lambda}$ may achieve the
limit $L^{-\alpha}$ where $\alpha=d$ in all regular i.e.,
non-critical points, while at the critical ones one can achieve
$\alpha>d$. For example in the class of one-dimensional systems studied in
\cite{za-co-gio,co-gio-za,co-ion-za} one has that the
estimation accuracy goes from order $L^{-1}$ in the regular
points to $L^{-2}$ at the critical points. 
{\bf iii)} At the critical point in the parameter space and for finite
temperature $T$ the covariance of the optimal estimator scales
like $T^{\beta}$. The exponent $\beta$ is related to the $\Delta_g$
above  and for a class QPTs is greater than zero \cite{ZCG}. In
this case by approaching zero temperature accuracy grows
unboundedly, {\em i.e} QFI diverges \cite{QLoss}. 
In contrast, at the regular points accuracy remains finite even
for $T\rightarrow 0$ (and finite system's size). In the
quasi-free fermionic case mentioned above one has $\beta=1$. 
\subsection{Finite-size corrections}\label{corrections}
In view of practical applications involving realistic samples, we
now consider corrections arising from the finite system size. 
As we have seen, when $L\to\infty$, the maximum of the QFI is located
at the critical point $\lambda_c$. Instead, for finite $L$, the
location of the maximum is shifted by an amount which goes to zero as
$L\to\infty$. To obtain an estimate of the shift one
must include off-scaling contribution. The previous formulae for the
scaling of $ds^2_B, \mathcal{H}$ in the off- and
quasi-critical regions can be combined in a single equation valid in a
broader regime:
\begin{eqnarray}
\frac{\mathcal{H}}{L^d}&=&L^{\Delta_{g}}\phi\left(z\right) +
L^{\Delta_{g}}D\left(\lambda\right) + \nonumber\\
& & L^{\Delta_{g}-\epsilon}C\left(\lambda\right) + \mbox{smaller terms}.   
\label{eq:superscaling} 
\end{eqnarray}
Here $z$ is the scaling variable $z=(L/\xi)^{1/\nu}$, $\epsilon>0$,
and $\phi$ is a 
scaling function satisfying $\phi(z)\sim z^{-\nu\Delta_{g}}$ when
$z\rightarrow\infty$ (the off-critical region), whereas  we must have
$\phi(0) \neq 0$ to comply with the behavior in the quasi-critical
region ($z\to 0 $). The existence of such a scaling function is a
consequence of the scaling hypothesis. Instead the functions
$C(\lambda), D(\lambda)$ are analytic around $\lambda_c$ and are
responsible for corrections to scaling. The maximum of $\mathcal{H}
(\lambda)$ defines a pseudo-critical point $\lambda^\ast_L$ whose
location, for large $L$ is given by
\begin{equation}
\lambda_{L}^{\ast}-\lambda_{c}\approx-\frac{\phi'}{\phi''}L^{-1/\nu}-\frac{D'}{\phi''}L^{-2/\nu}-\frac{C'}{\phi''}L^{-2/\nu-\epsilon}.
\end{equation}
One can say that the shift exponent of the QFI is, $1/\nu$, $2/\nu$,
or larger depending on the form of the functions above. 
A similar shift of the location of the quantum critical point may be
observed when the temperature is turned on.
\section{Optimal measurements}
\label{s:meas}
Our results establish the quantum limits to the degree of accuracy in 
estimating the coupling constant of a locally-interacting many-body 
system. It is crucial to stress again that the QCR bound is, in this 
case, attainable with the corresponding SLD representing an observable
to be measured in order to achieve the optimal estimation.  Of
course the SLD may be, in general a very complex observable which itself
depends on the ground or thermal state of the system.
For a family of pure states $\varrho (\lambda) = |\psi(\lambda)\rangle
\langle \psi(\lambda)|$, the SLD is easily derived as $$\LL_\mu(\lambda)=
2 \partial_\mu \varrho (\lambda)=2(|\psi\rangle\langle \partial_\mu\psi|
+|\partial_\mu\psi\rangle\langle \psi|)\:.$$
Using first order perturbation theory,
one finds $\langle n|\partial_\mu\psi\rangle=(E_n-E_0)^{-1} 
\langle n|\partial_\mu H|0\rangle$ where now the  $|n\rangle$'s ($E_n$'s) are
the eigenvectors (eigenvalues) of $H(\lambda)$ being $|0\rangle$ the ground state. 
Therefore, we may write the SLD as $$\LL_\mu (\lambda) = 2 [(P_0
\partial_\mu H G(E_0)+h.c.]$$ where $P_0=|0 \rangle\langle 0|$, $G(E)=P_1
[H(\lambda) - E]^{-1} P_1$, being $P_1 = I -P_0$. 
In the one parameter case the non-vanishing eigenvalues of the SLD
are given by $\pm 2\sqrt{\langle\partial\psi|\partial\psi\rangle 
-|\langle\psi|\partial\psi\rangle|^2}=\pm 2 ds_B/d\lambda$
from which the unknown parameter $\lambda$ can be estimated. 
The corresponding QFI is given by \cite{DG-qpt}
\begin{align}
{\cal H}_{\mu\nu} (\lambda) 
= 4 \sum_{n>0} \frac{\langle 0 | \partial_\mu H| n\rangle\langle n |
\partial_\nu H | 0\rangle}{(E_n-E_0)^2}\:.
\end{align}
From this expression one sees that the origin of the divergence of QFI
at the critical point is the vanishing (in the thermodynamical limit) of
one of the $E_n-E_0$ factors.
\subsection{\em Finite temperature}
Now we would like to show that also  classical i.e., temperature
driven, phase transitions provide in principle a valuable
resource for estimation theory. Consider the metric induced on
the (inverse) temperature axis by the mapping $\beta\rightarrow
\varrho(\beta).$ In \cite{DG-qpt} and \cite{ZCG} it has been shown
that
$ds^2 \sim d\beta^2 (\langle H^2\rangle_\beta
-\langle H\rangle_\beta^2)=d\beta^2 \beta^{-2} c_V(\beta)$, 
where $c_V(\beta)$ denotes the specific heat at the inverse
temperature $\beta.$ In \cite{DG-qpt} it was noticed that this
relation suggests a neat and deep interplay between Hilbert space
geometry and thermodynamics. Here we stress
that also quantum estimation gets involved.
Indeed, following the same lines used in the QPTs case,
one easily realizes that when the specific heat shows an anomalous
increase then the same happens to the QFI associated to the
parameter $\beta.$ From this fact stems that {\em at the
classical phase transitions with diverging specific heat one can
estimate temperature with arbitrarily high accuracy}. Conversely,
if the specific heat is bounded from above the estimation
accuracy of the temperature is bounded from below. Again, 
in analogy with the QPT case discussed above, if $\beta$
can be made dependent, in some known fashion,  on some other
parameter $\lambda^\prime$, then  this latter can be estimated
with better accuracy at the phase transition. In other
words, we have a quantitative statement of the intuitive
expectation about the fact that {\em thermometers} are more
precise at the points where changes of states of matter occur.
\subsection{Quantum discrimination}
In our analysis the problem of interest is that of estimating the
value of a parameter or a set of parameters. On the other hand,
when one deals with {\em discrimination} of parameters rather
than estimation, the relevant metric $ds^2_{QCB}$ can be derived
from the so-called quantum Chernoff bound \cite{qcb}, which arises
in the problem of discriminating two quantum states in the
setting of asymptotically many copies.  In view of the inequality
$\frac12 ds^2_B \leq ds^2_{QCB} \leq ds^2_B$ all the results
obtained here can be generalized to $ds^2_{QCB}$.  This means that
{\em quantum criticality is a resource for quantum
discrimination} as well.
\subsection{Applications: two-stage adaptive measurements}
A question may arise on how our results may be exploited in practice,
being the form of the SLD, which maximizes the Fisher
information, typically dependent on the true, unknown, state of the
quantum system. The apparent loophole in the argument is closed by
noticing that one can still achieve the same rate of
distinguishability by a two-stage adaptive measurement procedure
\cite{gill00}. Roughly speaking the estimation scheme goes as follows: 
one starts by performing a generic, perhaps suboptimal, measurement 
on a vanishing fraction of the copies of the system and obtains a 
preliminary estimation. Then one measures the remaining copies, taking 
the the preliminary estimation as the true value. This guarantees that 
the Fisher information obtained by the second series of measurements 
approaches the QFI as the number of measurements goes to infinity, and 
that the resulting estimator saturate the QCR bound.
Notice that the achievability of the QCR bound is 
ensured when a single parameter has to be estimated \cite{fuj06}, 
though the actual implementation of the optimal measurement,
which is general not unique \cite{sar06}, may be more or less
challenging depending on the specific features of the system 
under investigation. 
We also notice that any maximum-likelihood (ML) \cite{zhl}
estimator, defined as the estimator $\hat\lambda = \hat\lambda (x_1,
x_2, ...)$ maximizing the likelihood function ${\cal L}(\lambda) =
\prod_k^M p(x_k|\lambda)$, $M$ being the number of measurements and
$p(x|\lambda)$ the conditional probability density of the measured
quantity, is {\em consistent}, i.e. it converges in probability to the
true value and asymptotically ($M\gg 1$) {\em efficient}, i.e.  it
saturates the Cramer-Rao bound in the limit of many measurements, thus
achieving the ultimate bound in precision. 
\section{Quantum estimation in the BCS model}
\label{s:ebcs}
In order to appreciate the critical enhancement of precision 
in a specific physical situation we consider a fermionic tight-binding
model with pair creation {\em i.e.} the BCS-like model defined by the
Hamiltonian
\begin{align}
H_J  =  &-J\sum_{i=1}^{L}\left(
c_{i}^{\dagger}c_{i+1}
+\gamma c_{i}^{\dagger}c_{i+1}^{\dagger}+
\hc\right) 
-2h\sum_{i=1}^{L}c_{i}^{\dagger}c_{i}
 \end{align}
where $L$ denotes the system size and periodic boundary conditions
have been used.  
Upon considering the thermal or the ground state of the above
Hamiltonian we have a uniparametric statistical model where $J$ is the parameter to
be estimated, $h$ is the external 
(tunable) field and the anisotropy $\gamma$ is a fixed quantity. For
$\gamma\neq0$ this model undergoes a quantum phase transition of
Ising type at $h=J$ (because of symmetries there is an analogous
critical point at $h=-J$). Precisely around the Ising transition point 
point we will investigate the enhancement in precision offered by
criticality. 
\par
Moving to Fourier space the Hamiltonian may be rewritten as
\begin{align}
H_J & = \sum_{k\in BZ}\epsilon_{k}\left(n_{k}+n_{-k}-1\right)+
\sum_{k\in
BZ}\left(i\Delta_{k}c_{k}^{\dagger}c_{-k}^{\dagger}+\hc\right)\nonumber\\
 & = \sum_{k>0}\left[-\epsilon_{k}\tau_{k}^{z}+\Delta_{k}\tau_{k}^{y}\right]
\label{blockH} 
 \end{align}
where $\epsilon_{k}=-J\cos\left(k\right)-h$, 
$\Delta_{k}=-J\gamma\sin\left(k\right)$, the Brillouin zone $BZ$ ranges
from $-\pi$ to $\pi$ and the momenta are of the form $k=2 \pi n/L$, with
integer $n$.
The expression (\ref{blockH}) for $H_J$   
follows from the observation that  
in the subspace spanned by $\left\{|0\rangle,c^\dag_k c^\dag_{-k}|0\rangle\right\}$ 
the operators $n_{k}+n_{-k}-1 = -\tau_{k}^{z}$ and
$\left(ic_{k}^{\dagger}c_{-k}^{\dagger}+\hc\right)  = \tau_{k}^{y}$,
represent a set of Pauli operators $\tau_k^j$ (and they are zero in the
complementary space).
\par
In the above quasi-spin formulation the Hamiltonian in Eq. (\ref{blockH}) has 
a block form, 
and so is the thermal state $\varrho_J=Z^{-1}
\exp\{-\beta H_J\}$. Being each block essentially two-dimensional, the
SLD of the model can be obtained by computing the SLD in each block,
with the aid of formula (18) of \cite{1DI}. One then arrives at
$$\LL (J)=\sum_{k>0}b_{k}^{z}\tau_{k}^{z}+b_{k}^{y}\tau_{k}^{y}\:,$$
where, at $T=0$ 
\begin{align}
b_{k}^{y}  = \frac{h}J \frac{\Delta_k \epsilon_k}{\Lambda_k^3} 
\qquad 
b_{k}^{z}  = \frac{h}J \frac{\Delta_k^2}{\Lambda_k^3} 
\end{align}
where $\Lambda_{k}=\sqrt{\epsilon_{k}^{2}+\Delta_{k}^{2}}$.
\par
Going back to real space we have 
\begin{align}
\LL(J)& =  \frac12 Lb^{z}\left(0\right)-\sum_{l,j}c_{l}^{\dagger}b^{z}
\left(l-j\right)c_{j} \nonumber \\ 
& +\frac12 \left[i\sum_{l,j}c_{l}^{\dagger}b^{y}
\left(j-l\right)c_{j}^{\dagger}+\hc\right]\,,\end{align}
where 
\[b^{z}\left(d\right)\equiv\frac{1}{L}\sum_{k\in BZ}e^{-ikd}b_{k}^{z}\]
and analogously for $b^{y}(d)$. For example, for $L=4$ the general
formula reduces to
\begin{align}
\LL(J)& = \frac{h \gamma}{(h^2+\gamma^2 J^2)^{3/2}} \left\{
J \gamma - \frac{J \gamma N}2 \right. \\  
&\left.+ \frac{J\gamma}2 \left[c_1^\dag c_3 +
c_2^\dag c_4 + \hc \right]+ h \left[ c_1^\dag c_2^\dag - c_1^\dag
c_4^\dag + \hc\right]
\right\}\nonumber 
\end{align}
which represents a collective measurement on the system. $N=\sum_j
c_j^\dag c_j$ is the total number operator.
\par
In general the coefficients $b^{z}(d)$ and $b^{y}(d)$ decay
exponentially at normal points of the phase diagram and thus the SLD is
a local operator. On the other hand, the decay is only algebraic in the
critical region $|h-J| L \lesssim 1$ and this corresponds to a SLD given
by a collective measurement. Indeed, at the Ising transition one can
show that $b^{y}(d) \sim d^{-1}$. For $b^{z}(d)$ the integral of Fourier
coefficient does not converge for large $L$, so that one has to keep the
sum with $L$ finite. The sum is well approximated by $b^{z}(d)\simeq
2.9\left|\frac{d}{L}-\frac{1}{2}\right| \left(-1\right)^{d}$.
Notice that for $\gamma=0$ the SLD at zero temperature is
identically zero since for any finite $L$ changing $J$ only
results in a level crossing. 
\par
When the temperature is turned on, one can draw similar conclusions. 
The local character of the optimal measurement is enhanced at regular
points $\forall L$ whereas in the critical regime nonlocal measurements
are needed. More specifically, the coefficients $b^{z,y}(d)$ decay 
exponentially at regular points and algebraically in the region $|h-J|
\lesssim T$.
\par 
The ultimate precision is determined by the QFI  ${\cal H}_J$, which,
for such quasi-free Fermi system reduces to (see \cite{DG-qpt})
\begin{align}
\mathcal{H}(J)=\sum_{k>0}\left(\partial_{J}\vartheta_k\right)^{2}
=\sum_{k>0}\frac{h^{2}\gamma^{2}\sin\left(k\right)^{2}}{\Lambda_{k}^{4}}
\end{align}
where $\vartheta_k = \hbox{arctan}\left(\epsilon_k/\Delta_k\right)$.
Upon introducing the scaling variable $z=L\left(h-J\right)$ one is
interested in the behavior around the Ising transition, {\em i.e.} 
for $z\ll 1$ where one can observe superextensive behavior.
Indeed, upon expanding for small $z$ and using Euler-Maclaurin
formula, one finally gets
\begin{align}
\mathcal{H}(J) & =  \frac{L^{2}}{24J^{2}\gamma^{2}}-\frac{L}{2
\pi^{2}J^{2}\gamma^{2}}\nonumber \\  & +  \frac{z}{L}\left[
\frac{\left(\gamma^{2}-1\right)L^{2}}{12J^{3}\gamma^{4}}+ 
O\left(L\right)\right] \nonumber \\
 & -  \frac{z^{2}}{L^{2}}\left[\frac{L^{4}}{720J^{4}\gamma^{4}}
 +O\left(L^{2}\right)\right]+O\left(z^{3}\right)\:. 
 \end{align}
From this expression one can explicitly see the improvement of precision
due to the underlying quantum critical behavior of the system.
Besides, one can locate the pseudocritical point, defined as 
value of the field leading to the maximum of
$\mathcal{H}(J)$. Differentiating the above formula with respect to
$h$ one obtains 
$$h_{L}^{\ast}=J+30J\left(\gamma^{2}-1\right)L^{-2}+O\left(L^{-3}\right)\:,$$
and use this information to achieve the quantum Cramer-Rao bound 
also for finite-size systems.
\par
With the notations of section \ref{corrections}, we note we can
write the scaling form of 
$\mathcal{H}(J)/L$ as in Eq.~(\ref{eq:superscaling})
with $\Delta_{g}=\nu=\epsilon=1$ and 
\begin{align}
\phi\left(z\right) & = \frac{1}{24J^{2}\gamma^{2}}-\frac{z^{2}}{720J^{4}\gamma^{4}}+O\left(z^{3}\right)\\
D\left(h\right) & = \frac{\left(\gamma^{2}
-1\right)}{12J^{3}\gamma^{4}}\left(h-J\right)+O\left[\left(h-J\right)^{2}\right]\\
C\left(h\right) & =
\frac{1}{2\pi^{2}J^{2}\gamma^{2}}O\left(h-J\right)\:.
\end{align}
Having $\phi'(0)=0$ the shift exponent turns out to be $2/\nu=2$
\par
At finite temperature and in the thermodynamical limit 
the QFI has been calculated previously for quasi-free Fermi
system \cite{ZCG}. In the present case we obtain
\begin{align}
\frac{\mathcal{H}(J)}{L} = & \frac{\beta^{2}}{8\pi}
\int_{0}^{\pi}\frac{dk}{\cosh^{2}\left(\beta\Lambda_{k}/2\right)}
\frac{\left(J+h\cos\left(k\right)\right)^{2}}{\Lambda_{k}^{2}}\\
+ & \frac{1}{2\pi}\int_{0}^{\pi}\frac{\cosh\left(\beta
\Lambda_{k}\right)-1}{\cosh\left(\beta\Lambda_{k}\right)}
\frac{h^{2}\gamma^{2}\sin\left(k\right)^{2}}{\Lambda_{k}^{4}}dk\:.
\end{align}
In this case we verified numerically that the maximum 
of $\mathcal{H}(J)$ always occurs at $h=J$ where it 
has a cusp. Then, since when the temperature goes to zero 
the first integral vanishes we evaluate the second term at 
the critical point. For $h=J$ the dispersion $\Lambda_{k}$ 
is linear around $k=\pi$, which gives the dominant
contribution to the integral. One then obtains  
$$
\mathcal{H}(J)=\frac{2\mathcal{C}}{\pi^{2}}
\frac{L}{T\left|J\gamma\right|}+O\left(T^{0}\right)\:,$$
$\mathcal{C}=0.915$ being the Catalan constant.
From the above formula one can again appreciate the enhancement 
of the bound to precision occurring when $T$ goes to zero in the
critical regime.
\section{Conclusions and outlooks}
\label{s:out}
In conclusion, upon bringing together results from the geometric
theory of quantum estimation and the geometric theory of quantum
phase transition we have quantitatively shown that phase
transitions represent a resource for the estimation of
Hamiltonian parameters as well as of temperature.  To this aim we
used  the quantum Cramer-Rao bound and the equivalence of the
notion of quantum Fisher metric and that of quantum (ground or
thermal) state metric. We have also found an explicit form of the
observable achieving the ultimate precision.  A specific example involving  
a fermionic tight-binding model with pair creation 
has been presented in order to illustrate the critical enhancement
and the properties of the optimal measurement.
The improvement in estimation tasks
brought about by quantum criticality is reminiscent of the one
associated to quantum entanglement in computational as well as
metrological tasks.  The analysis reported in this paper  makes
an important point of principle and establishes the ultimate
quantum  limits to the precision with which one can estimate
coupling constants characterizing a quantum Hamiltonian. Since the
Hamiltonian completely characterize  the quantum dynamics one can
say that our results shed light on  the ultimate limits imposed
by quantum theory to the observer capability of knowing a quantum
dynamics.  Remarkably the boundaries between different phases of
quantum matter are where these limits gets looser and a deeper
knowledge can be achieved. 
\par $ $ \par
MGAP would like to acknowledge useful discussions with Alex Monras 
and Paulina Marian.

\end{document}